\documentclass[conference]{IEEEtran}
\IEEEoverridecommandlockouts

\makeatletter
\def\endthebibliography{%
  \def\@noitemerr{\@latex@warning{Empty `thebibliography' environment}}%
  \endlist
}
\makeatother

\usepackage{cite}
\usepackage{amsmath,amssymb,amsfonts}
\usepackage{graphicx}
\usepackage{textcomp}
\usepackage{xcolor}
\usepackage{subcaption}
\usepackage{tcolorbox}
\usepackage{xspace}
\usepackage{algcompatible}
\usepackage{multirow}
\usepackage[ruled,linesnumbered]{algorithm2e}

\tcbuselibrary{listings,breakable}

\def\BibTeX{{\rm B\kern-.05em{\sc i\kern-.025em b}\kern-.08em
    T\kern-.1667em\lower.7ex\hbox{E}\kern-.125emX}}

\newcommand{\ours}{HPDR\xspace}

\begin{document}

\newtcbox{\inlinebox}[1][]{enhanced,
 box align=base,
 nobeforeafter,
 colback=gray!40,
 colframe=white,
 size=small,
 left=0pt,
 right=0pt,
 boxsep=2pt,
 #1}

\title{\ours: High-Performance Portable Scientific Data Reduction Framework}

\author{
\IEEEauthorblockN{Jieyang Chen$^{*}$, Qian Gong$^{\diamond}$, Yanliang Li$^{*}$, Xin Liang$^{\dagger}$, Lipeng Wan$^{\ddagger}$,  Qing Liu$^{\bigtriangleup}$, Norbert Podhorszki$^{\diamond}$, Scott Klasky$^{\diamond}$}
\IEEEauthorblockA{
$^*$University of Oregon, OR, USA\\
$^\diamond$Oak Ridge National Laboratory, TN, USA\\
$^\dagger$University of Kentucky, KY, USA\\
$^\ddagger$Georgia State University, GA, USA\\
$^\bigtriangleup$New Jersey Institute of Technology, NJ, USA\\
jieyang@uoregon.edu, gongq@ornl.gov, leonli@uoregon.edu, xliang@uky.edu\\
lwan@gsu.edu, qing.liu@njit.edu, pnorbert@ornl.gov, klasky@ornl.gov
}

\thanks{This manuscript has been authored by UT-Battelle, LLC, under contract DE-AC05-00OR22725 with the US Department of Energy (DOE). The US government retains and the publisher, by accepting the work for publication, acknowledges that the US government retains a non-exclusive, paid-up, irrevocable, world-wide license to publish or reproduce the submitted manuscript version of this work, or allow others to do so, for US government purposes. DOE will provide public access to these results of federally sponsored research in accordance with the DOE Public Access Plan (http://energy.gov/downloads/doe-public-access-plan).}
}
\maketitle

\begin{abstract}
The rapid growth in scientific data generation is outpacing advancements in computing systems necessary for efficient storage, transfer, and analysis, particularly in the context of exascale computing. With the deployment of first-generation exascale computing systems and next-generation experimental facilities, this gap is widening and necessitates effective data reduction techniques to manage enormous data volumes.
Over the past decade, various data reduction methods, including lossless compression, error-controlled lossy compression, and data refactoring, have been developed to accelerate I/O in scientific workflows. Despite significant reductions in data volume, these methods introduce considerable computational overhead, which can become the new bottleneck in data processing. To mitigate this, GPU-accelerated data reduction algorithms have been introduced. However, challenges remain in their integration into exascale workflows, including limited portability across different GPU architectures, substantial memory transfer overhead, and reduced scalability on dense multi-GPU systems.
To address these challenges, we propose \ours, a high-performance and portable data reduction framework. \ours is designed to enable the execution of state-of-the-art reduction algorithms across diverse processor architectures while reducing memory transfer overhead to 2.3\% of the original, resulting in up to 3.5$\times$ faster throughput compared to existing solutions. It also achieves up to 96\% of the theoretical speedup in multi-GPU settings. 
In addition, evaluations on accelerating I/O operations at scale up to 1,024 nodes of the Frontier supercomputer demonstrate that \ours can achieve up to 103 TB/s reduction throughput, providing up to 4$\times$ acceleration in parallel I/O performance compared to existing data reduction routines. This work highlights the potential of \ours to significantly enhance data reduction efficiency in exascale computing environments.
\end{abstract}


\section{Introduction}
The generation of scientific data is outpacing the rate of improvement in the computing systems needed to save, transfer, and analyze data. 
With the deployment of exascale computing workflows that consist of the first generation of exascale computing systems~\cite{fnt, aur} and the next generation of experimental and observation facilities~\cite{iter, ska, golaz2019doe}, there will be an even wider compute and I/O gap in exascale computing of the foreseeable future.
For example, XGC simulation code~\cite{ku2009full, chang2004numerical}, which simulates a fusion reactor with magnetically confined plasma, can generate data at a rate of exabytes per day.
In addition, Square Kilometer Array~\cite{taylor2004science}, the world’s largest radio telescope, is anticipated to have a raw data rate of approximately 2 PB/s, or 600 PB/yr after hierarchical beam-forming reduction, all to be stored on the buffer file system~\cite{chrysostomou2020operating}.
These scientific workflows highlight the growing need to integrate data reduction in workflows to accelerate data storage, sharing, and analysis.

During the past decade, many data reduction techniques, such as lossless compression~\cite{knorr2021ndzip, liu2021improving, zhang2023gpulz, shah2023lightweight, mao2022trace}, error-controlled lossy compression~\cite{zhao2020significantly, zhao2021optimizing, gong2023mgard, liang2021mgard+, lindstrom2014fixed, gong2024general, banerjee2023fast, banerjee2023online}, and data refactoring~\cite{liang2021error, chen2021accelerating, wu2024error}, have been developed to accelerate I/O in scientific workflows. 
While they provide significant data volume reduction, modern data reduction routines also bring a considerable amount of computational overhead, which tends to increase as more advanced reduction algorithms are built.
Since such reduction overhead may become the new bottleneck of I/O, several Graphics Processing Unit (GPU) accelerated data reduction algorithms have recently been developed.
For instance, nvCOMP~\cite{nvcomp} offers a collection of GPU lossless compression developed by NVIDIA. 
Lossy compression such as MGARD~\cite{gong2023mgard}, SZ~\cite{zhao2021optimizing}, and ZFP~\cite{lindstromzfp} also developed GPU accelerated parallel design: MGARD-GPU~\cite{gong2023mgard}, cuSZ~\cite{tian2020cusz, tian2021cusz, huang2023cuszp}, and ZFP-CUDA~\cite{lindstromzfp}.

Although GPU-accelerated data reduction offers performance acceleration compared with their serial implementation, scientists still face several significant challenges in using them to accelerate I/O in exascale computing workflows: (1) Current data reduction has limited portability across architectures. Most existing reduction algorithms are accelerated for Nvidia GPUs with very limited support on other GPU architectures (such as AMD and Intel GPUs) that are used in current exascale computing systems. 
Without portability, data cannot be easily shared across computing facilities.
(2) A large gap exists between the performance obtained on GPU and the perceived performance by user applications. This is because most reduction algorithms are memory-bound, but the overhead brought by memory operations such as management and transfer are often overlooked. 
Those operations can create a large overhead or dominate the reduction pipeline.
(3) Modern large-scale computing systems often use dense GPU architecture that packs multiple GPUs into one computing node.
While such types of architecture bring performance advantages, they also raise challenges for scalability for memory-bound operations, which is typically the case in data reduction. 
This is because all GPUs share the same runtime, where memory operations issued concurrently can easily cause contention that degrades the scalability of data reduction routines. 
To overcome those challenges and help modern scientific workflows benefit from data reduction, we design \ours. 
\ours is a high-performance and portable framework tailored for efficiently running reduction algorithms across different processor architectures.
In addition, our framework is highly extensible to execute new reduction algorithms and support new processor architectures via programming models (e.g., CUDA~\cite{luebke2008cuda}, HIP~\cite{amd_hip}, etc.) and general-propose portability libraries (i.e., SYCL~\cite{reyes2016sycl}, Kokkos~\cite{edwards2014kokkos}, etc.).

Our paper makes the following contributions.
\begin{itemize}
    \item We propose a portable data reduction framework, \ours, featuring a series of runtime abstractions to optimize the performance and scalability of common data reduction algorithms across multi-GPUs and multi-core architecture. We demonstrate how three state-of-the-art data reduction tools---MGARD, ZFP, and Huffman encoding---can be implemented using \ours, ensuring compatibility across five distinct CPU and GPU architectures. 
    \item We identify CPU-GPU memory transfer as a significant performance bottleneck, which is often overlooked in existing state-of-the-art data reduction works. Furthermore, for applications that continuously generate data, reduction and data movement must be optimized in tandem to maximize overall performance. To address this, we propose a highly optimized pipeline. By adaptively overlapping reduction with data transfer and coordinating the execution of multiple asynchronous tasks, \ours reduces the data transfer overhead to 2.3\% of the original and accelerates end-to-end reduction throughput by up to $3.5\times$ for the three state-of-the-art reduction pipelines. When scaling on multiple GPUs, reduction optimized with \ours achieves up to 96\% of the theoretical speedup, compared to a mere 74\% with non-optimized designs. 
    \item To demonstrate the I/O acceleration using \ours for reading and writing scientific data at scale, we integrate \ours with ADIOS2 I/O library~\cite{godoy2020adios} on two leadership-class supercomputers, Summit and Frontier, at Oak Ridge Leadership Computing Facility (OLCF) with scale up to 1,024 nodes. Our evaluation shows that \ours achieves up to 103 TB/s reduction throughput and provides up to $4\times$ parallel I/O acceleration compared with I/O with existing data reduction routines.
\end{itemize}

The rest of the paper is organized as follows. Section~\ref{sec: background} introduce backgrounds and challenges of achieving high-performance data reduction on modern computing systems. Then, we introduce our novel high-performance portable data reduction framework in Section~\ref{sec:mgard-x} and demonstrate the applicability of our reduction framework for three state-of-the-art reduction pipelines in Section~\ref{sec:imp}. Moreover, to optimize reduction process on accelerators we further introduce an optimized reduction pipeline design in Section~\ref{sec:pipeline-opt}. Finally, we provide a thorough experimental evaluation of our proposed works in Section~\ref{sec:eval}.
\section{Background and Problem Statement}
\label{sec: background}
\subsection{Scientific data reduction}
In the past decade, various data reduction techniques have been developed to reduce the volume of scientific data since they has long been considered as a powerful tool for alleviating I/O and storage costs. 
Among those, error-bounded lossy compression techniques have gained popularity.
Methods like MGARD~\cite{gong2023mgard}, SZ~\cite{zhao2021optimizing}, ZFP~\cite{lindstromzfp}, and ISABELA~\cite{lakshminarasimhan2013isabela}, offer remarkable compressibility while guaranteeing that the difference between original and the reconstructed data remains within user prescribed error bounds. Most error-bounded lossy compressors follows a three-step processes, involving decorrelation, quantization, and lossless encoding, supported by mathematical theories to ensure error-bound satisfaction. Specifically, the decorrelation and quantization first transform original data into a set of discrete coefficients with lower entropy, enabling efficient encoding. Then, lossless encoding compress the coefficients to achieve data reduction. 

Although modern data reduction techniques can greatly reduce data volume, the computational costs of reduction can sometimes be non-trivial, leading to situations where the time spent on reduction and reconstruction outweighs the benefits gained from reduced data read and write time~\cite{liang2021error}. To minimize the data reduction overhead, many data reduction have been re-designed for parallel execution on CPU or GPU accelerators~\cite{chen2021accelerating, tian2021cusz}.






\subsection{Challenges of achieving high-performance data reduction}
Despite many data reduction optimizations that have been done to improve their performance on various types of processors, there are two major overlooked challenges that hinder application users from obtaining low-cost data reduction. 

\underline{Diverse processor architectures:} 
Modern computing systems are equipped with diverse types of processors, such as CPUs and GPUs, designed with different micro-architectures.
Because of their fundamental difference in terms of architecture, a data reduction pipeline typically leverages different algorithm designs on different processor types.
Although different algorithm designs may share the same foundational theory and method, their difference can cause data portability challenges: data reduced by one type of processor cannot be reconstructed by another type of processor with a guarantee.
Because of such portability challenges, application users are forced to use the most compatible processor to make sure application data are (1) shareable across different components workflows that may reside on different systems; and (2) retrievable on future systems that use new architectures.  
However, the most compatible processors, such as single-core CPUs, cannot guarantee the best reduction performance.

\begin{figure}[h!]
    \centering
    \includegraphics[width=0.45\textwidth]{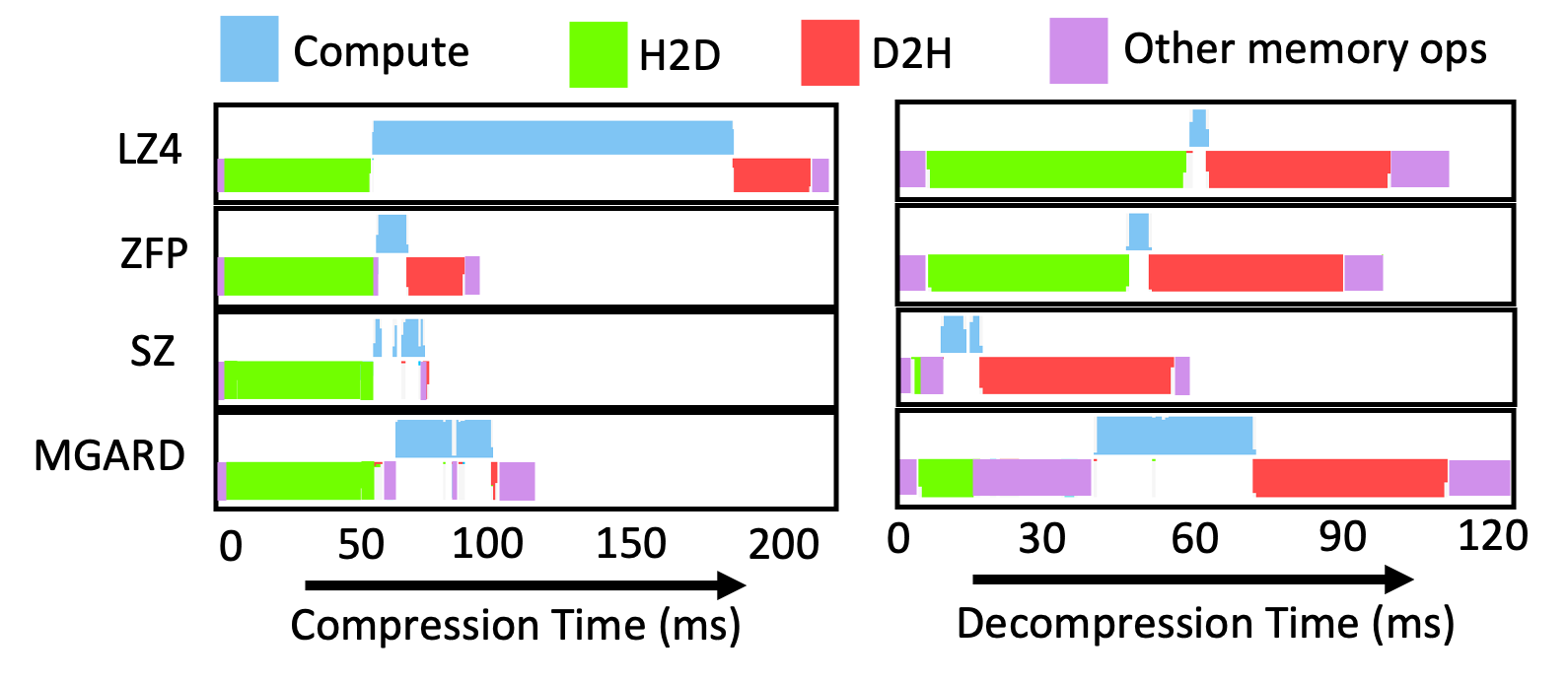}
    \caption{Time breakdown of reducing a 500 MB NYX data~\cite{almgren2013nyx} using four different reduction pipelines on a V100 GPU. $1e^{-2}$ error bound is used for lossy compression. Both application and I/O buffers are on the host.}
    \label{org-timeline}
\end{figure}
\underline{Large gap between GPU kernel and end-to-end performance:}
Although many works~\cite{chen2021accelerating, huang2023cuszp, lindstromzfp, zhang2023fz, flint2024using, tian2021revisiting} have been done to optimize the throughput of data reduction on GPU,
the improvement can hardly be transformed into the advancement of end-to-end performance perceived by application users.
This is because the latency of data transfers is often overlooked.
Those include memory copy between application buffer, reduction buffer, and I/O buffer.
Such data transfers cannot be easily eliminated as host memory is typically used by applications to save output data before reduction and/or I/O libraries as internal serialization buffers before I/O operations.
Unlike scientific applications where computation dominates costs, data reduction algorithms are typically memory bound~\cite{chen2021accelerating}, so the cost of memory operations brings considerable performance overhead.
Figure~\ref{org-timeline} shows the profiling results of four state-of-the-art GPU compression pipelines. We can see a large amount of time is spent on memory operations.
Specifically, 34 - 89\% of the total time is spent on memory operations (e.g., H2D and D2H) during compression or decompression pipeline. 
This ratio will increase if reduction kernels are further accelerated.
\section{\ours portable data reduction framework}
\label{sec:mgard-x}
\begin{figure}[ht]
\centering
\includegraphics[width=0.5\textwidth]{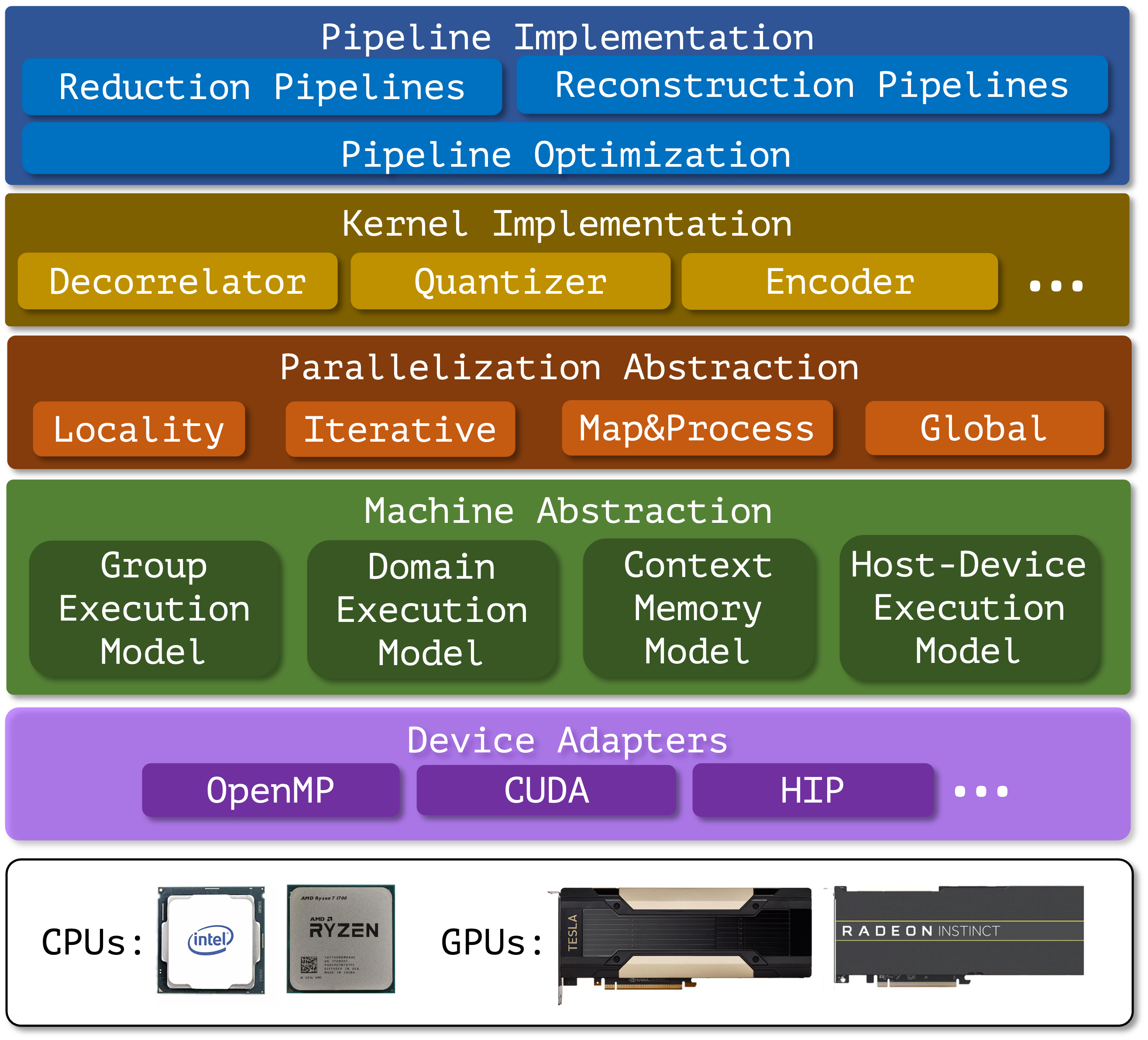}
\caption{High-perf. portable data reduction framework (\ours)}
\label{framework}
\vspace*{-1em}
\end{figure}
In this work, we propose a novel high-performance portable data reduction framework aiming to streamline data sharing across systems with different processor architectures. 
Figure~\ref{framework} shows the overall framework structure of \ours, including five software architectural layers. All layers work together to enable portability and optimizations that allow data reduction algorithms to take advantage of the best processor on a system, while producing portable data that can be reconstructed on another system.
In the rest of this section, we discuss three bottom layers: parallel abstraction, machine abstraction, and device adapter.
We will discuss how to implement data reduction algorithms on top of the three layers in Section~\ref{sec:imp} and pipeline optimization in Section~\ref{sec:pipeline-opt}.

\begin{figure*}[h!]
\centering
\includegraphics[width=1\textwidth]{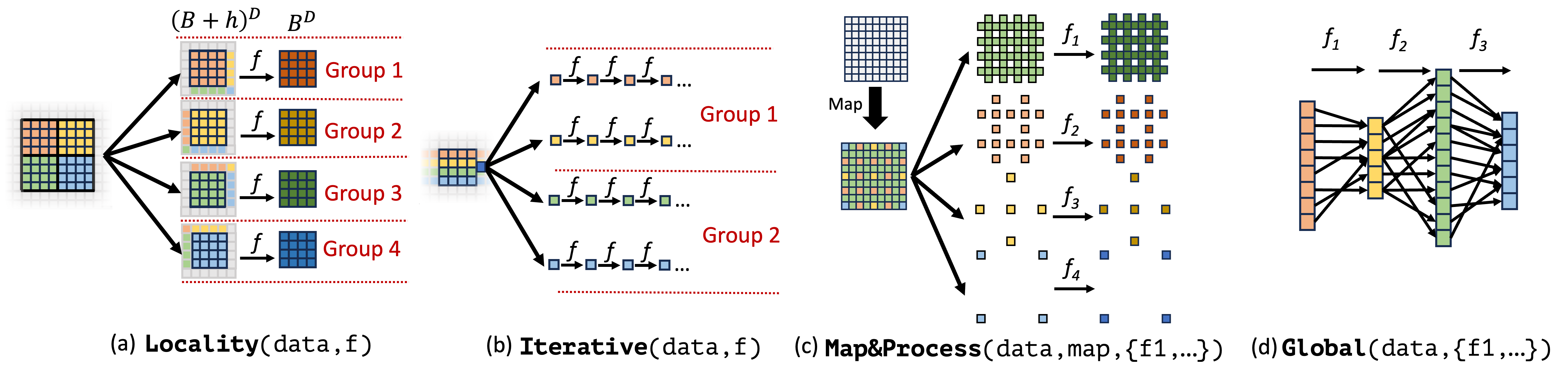}
\vspace*{-1em}
\caption{Parallel Abstractions in \ours}
\label{parallel-abstraction}
\vspace*{-1em}
\end{figure*}

\subsection{Parallelization Abstraction}
The parallelization abstraction layer enables data reduction algorithms to express their fine-grain parallelisms effectively.
Specifically, we introduce four example parallelization abstractions currently built in \ours.
\subsubsection{Locality Abstraction}
Many reduction algorithms exploit the correlations in data across multiple dimensions. 
For instance, MGARD uses a multi-dimensional multi-level decomposition. ZFP applies a block-wise compression on every $4^{dim.}$ cubic of data.
Figure~\ref{parallel-abstraction}(a) shows the locality abstraction designed for exploiting decorrelating the data using an algorithm-defined function $f$.
In this abstraction, the input data is parallelized by decomposing into blocks with customizable sizes and halo regions.
Then, a group of threads cooperatively executes $f$ on each block.
In addition, blocks can be loaded into a faster memory tier throughout the entire processing pipeline. 

\subsubsection{Iterative Abstraction} Some algorithms also need to process data iteratively along one dimension by repeatedly executing a function $f$. For example, solving tri-diagonal systems in MGARD. Figure~\ref{parallel-abstraction}(b) shows that iterative abstraction parallelizes the processing of each vector of data among different threads with every $B$ vector organized into a group. Similar to locality abstraction, each group of vectors is cached in a faster memory tier for better performance.

\subsubsection{Map And Process Abstraction} For decomposed-based data reduction such as MGARD, data needs to be decomposed into a hierarchy, and each level of the hierarchy needs to be processed individually. Figure~\ref{parallel-abstraction}(c) shows the map and process abstraction. The input data is first mapped to the subsets, and each is processed using a different function. 

\subsubsection{Global Pipeline Abstraction}
Many encoding algorithms, such as Huffman encoding, exploit the correlation and redundancy in the whole data domain. 
Also, parallel serialization operation commonly used in many compression algorithms also needs to collect global information to compact bit streams.   
This requires global-wide processing that allows all threads to collaborate globally. Figure~\ref{parallel-abstraction}(d) shows the global parallel abstraction. All data is processed at the same time with global synchronizations to enable communication and data exchange among all threads.


\begin{figure}[h!]
\centering
\includegraphics[width=0.5\textwidth]{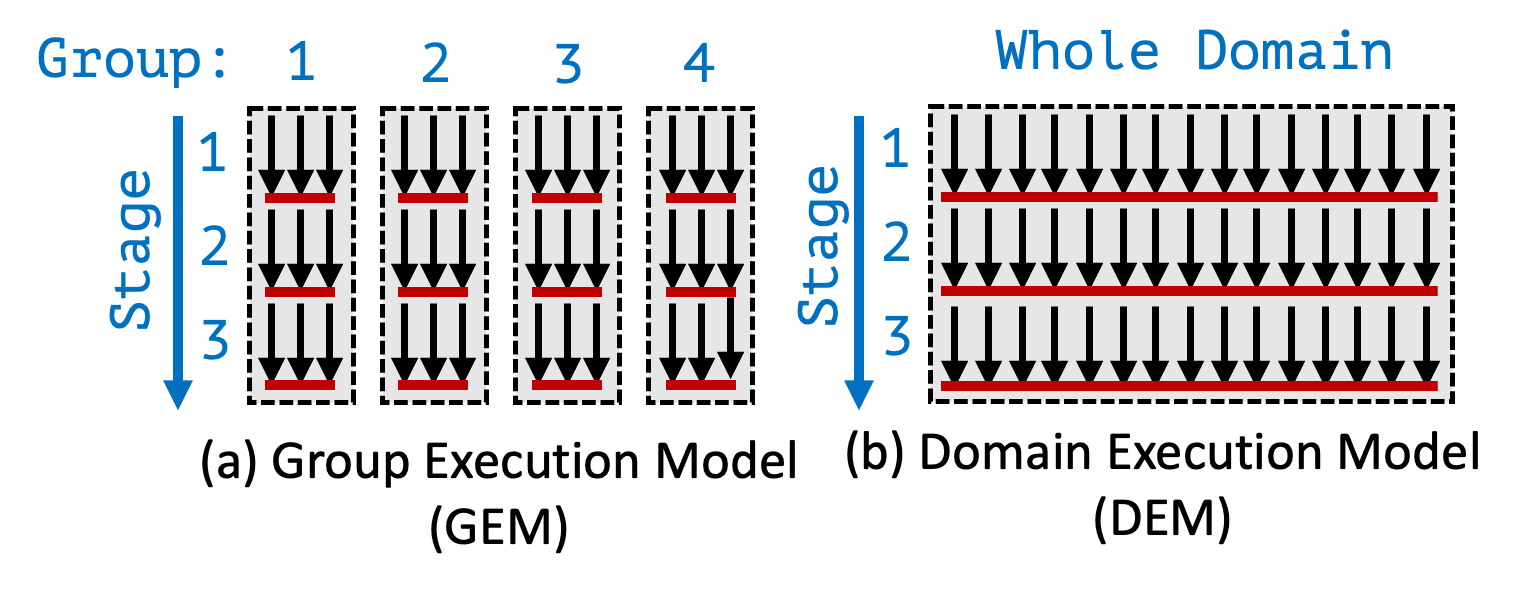}
\vspace*{-2em}
\caption{Group and Domain Execution Models}
\label{pm}
\vspace*{-1em}
\end{figure}

\begin{table}[ht]
\centering
\caption{Mapping Parallel Abstractions to Execution Models}
\label{abs-em}
\begin{tabular}{|c|c|c|}
\hline
Parallel Abstraction             & \textbf{GEM}            & \textbf{DEM}         \\ \hline
\textbf{Locality}     & Block$\rightarrow$Group     & -            \\ \hline
\textbf{Iterative}    & B*Vectors$\rightarrow$Group & -            \\ \hline
\textbf{Map\&Process} & -    & All Subsets$\rightarrow$Whole Domain            \\ \hline
\textbf{Global}       & -               & Domain$\rightarrow$Whole Domain \\ \hline
\end{tabular}
\end{table}

\subsection{Machine Abstraction} To execute the four parallel abstractions, we design three execution models and one memory management model to abstract the underlying hardware: the Group Execution Model (GEM), Domain Execution Model (DEM), Host-Device Execution Model (HDEM), Context Memory Model (CMM).
We will introduce GEM, DEM, and CMM in this section and disucss HDEM in Section~\ref{sec:pipeline-opt}.
GEM partitions threads into groups and executes independently. DEM, put all threads in one domain and execute in a synchronized fashion.
Both DEM and GEM support multi-stage execution such that multiple operations sharing the same execution model can be fused into one model for more efficient execution.
Table~\ref{abs-em} shows how the four parallel abstractions are mapped to execution models. 
For locality abstraction, blocks are 1:1 mapped to groups in GEM to exploit coarse-grain parallelism across blocks and fine-grain parallelism within a block.
For iterative abstraction, vectors are B:1 mapped to groups in GEM so that we can exploit memory locality across vectors.
Map\&process and global pipeline are mapped to DEM such that the whole data domain can be processed at the same time.


In addition to GEM and DEM, we build a memory management model, CMM, to manage the context memory of the reduction process.
One major performance bottleneck in data reduction that is often overlooked is the overhead of memory allocation management.
Data reduction algorithms tend to have low arithmetic intensity, so the cost of memory-related operations such as allocations for the purpose of building reduction context can easily dominate the cost of a reduction pipeline (e.g., other memory operations as shown in Figure~\ref{org-timeline}).
High memory allocation overhead can negatively impact end-to-end I/O performance greatly when a reduction pipeline is repeatedly invoked by scientific applications (e.g., every write iteration).
In addition, when a data reduction pipeline is deployed on a computing system with multiple GPUs that share the same runtime system (e.g., a multi-GPU compute node), all memory allocation management operations typically cannot be efficiently parallelized due to shared internal memory management, which can greatly impact the scalability of data reduction. To overcome such a challenge, we propose a context memory management optimization that uses a hash map to cache reduction contexts so they can be reused across repetitive reduction processes that share similar data characteristics. With this optimization, all memory allocations associated with a context will be persistent across calls to minimize the need for allocations.

\begin{table}[]
\caption{Mapping Execution Models to Devices}
\label{em-dev}
\begin{tabular}{|c|cc|c|c|c|}
\hline
Models               & \multicolumn{2}{c|}{Res. Mapping}                      & OMP       & CUDA      & HIP     \\ \hline
\multirow{4}{*}{GEM} & \multicolumn{1}{c|}{\multirow{2}{*}{Compute}} & Thread & Serial     & Core      & Core      \\ \cline{3-6} 
                     & \multicolumn{1}{c|}{}                         & Group  & Core      & SM        & CU      \\ \cline{2-6} 
                     & \multicolumn{1}{c|}{\multirow{2}{*}{Staging}} & Data   & Cache     & ShMem.    & ShMem.  \\ \cline{3-6} 
                     & \multicolumn{1}{c|}{}                         & Order  & Serial     & Blk. Sync.     & Blk. Sync.   \\ \hline
\multirow{3}{*}{DEM} & \multicolumn{1}{c|}{Compute}                  & Domain & All Cores & All Cores & All SUs \\ \cline{2-6} 
                     & \multicolumn{1}{c|}{\multirow{2}{*}{Staging}} & Data   & DRAM      & DRAM      & DRAM    \\ \cline{3-6} 
                     & \multicolumn{1}{c|}{}                         & Order  & Serial     & Grid Sync.      & Grid Sync.    \\ \hline
\end{tabular}
\end{table}

\subsection{Device Adapters} 
To enable portability support for data reduction, device adapters are designed to execute the two execution models on different architectures in a compatible way.
Device adapters have two major design advantages: (1) Separately designed device adapters enable us to fully exploit the functionalities that is only available for certain hardware architecture or programming models.
For example, whole domain synchronization can be more efficiently enabled using Cooperative Groups exclusively available in CUDA and HIP.
(2) \ours can be easily extended to support newer architectures or leveraging general-propose portability libraries such as Kokkos~\cite{trott2021kokkos} and SYCL~\cite{alpay2020sycl} by implementing new device adapters. 
Currently, we implement three device adapters: OpenMP, CUDA, and HIP for multi-core CPUs, Nvidia GPUs, and AMD GPUs.
Table~\ref{em-dev} shows how the two execution models are mapped to device adapters.
For OpenMP, GEM is executed by parallelizing groups among CPU cores, and the workload of each group is executed sequentially. Such a parallelization strategy can improve memory efficiency by letting each CPU core exploit data locality within a group. When executing a multi-stage GEM model, the execution order is maintained through sequential execution, and working data can be staged in the cache for sharing between stages. For DEM, OpenMP parallelizes all threads in the domain along all CPU cores to maximize parallelism. When there is more than one stage, the execution order is maintained through sequential execution, and working data among stages are shared via DRAM as it might be too large to fit into the cache.
For CUDA and HIP, GEM is executed by parallelizing groups among Streaming Multiprocessors (SMs) for CUDA or Compute Units (CUs) for HIP, and the workload of each group is parallelized on GPU cores. When executing a multi-stage GEM model, the execution order is maintained through threadblock-level synchronization, and working data can be staged in shared memory for sharing between stages. For DEM, all threads in the domain are parallelized along all GPU cores to maximize parallelism. When there is more than one stage, the execution order is maintained through Cooperative Groups, and working data among stages are shared via DRAM.

\section{Data Reduction Pipeline Case Study}
\label{sec:imp}
To demonstrate the applicability of \ours, we show how to build three data reduction pipelines: MGARD compression~\cite{gong2023mgard}, 
ZFP fixed-rate
data reduction~\cite{lindstromzfp}, and Huffman lossless compression~\cite{tian2021revisiting}.

\SetKwInOut{KwInOut}{In/Out}
\SetKwInOut{KwIn}{In}
\SetKwInOut{KwOut}{Out}
\begin{algorithm}[ht!]
\caption{MGARD Lossy Compression}
\label{mgard-imp}
\SetKwFunction{FMain}{MGARD}
\SetKwProg{Fn}{Function}{:}{\KwRet{$u_{compressed}$}}
\Fn{\FMain{}}{
\KwIn{Data $u$}
\KwIn{Error bound $e$}
$hierarchy \leftarrow u$\\
$mc = \{\}$\\
$l \leftarrow 0$\\
\While{$l < hierarchy.total\_levels$} {
$mc_{l} \leftarrow$ \texttt{Locality(u, lerp())}\\
$correction_{l} \leftarrow mc_{l}$\\
\texttt{Locality($correction_{l}$, mass\_trans())}\\
\texttt{Iterative($correction_{l}$, tridiag())}\\
\texttt{Locality(u, add($correction_{l}$))}\\
$mc =  mc \cup mc_{l}$\\
$l \leftarrow l + 1$ \\
}
$mc_{quantized} \leftarrow $ \texttt{MapAndProcess(mc, hierarchy, quantization(e))}\\
$u_{compressed} \leftarrow $ \texttt{Huffman($mc_{quantized}$)}\\
}
\end{algorithm}
\subsection{MGARD lossy compression}
\begin{figure}[h!]
    \centering
    \includegraphics[width=0.5\textwidth]{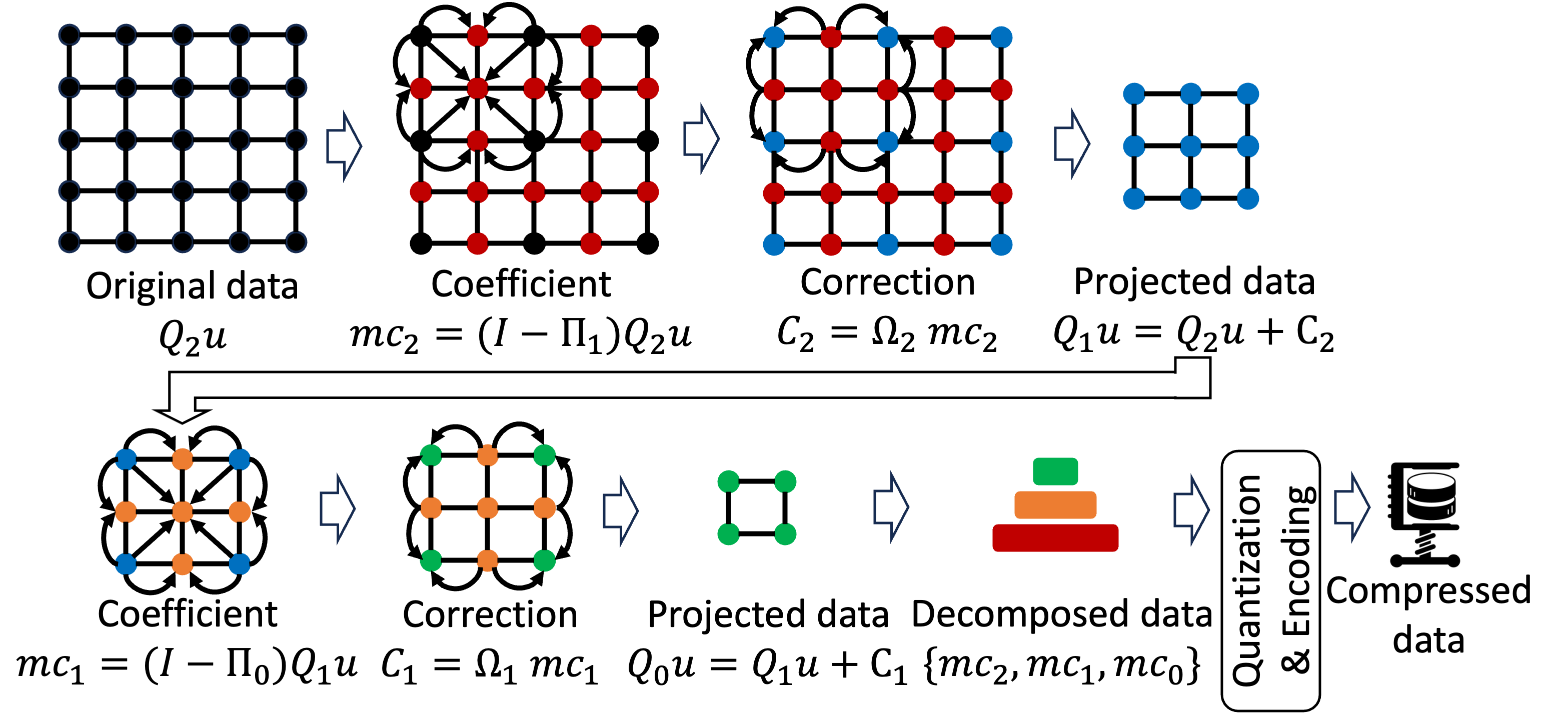}
    \caption{MGARD compression pipeline}
    \label{mgard-pipeline}
    \vspace*{-1em}
\end{figure}
MGARD is designed to compress both uniform and non-uniform grids while preserving the error of Quantities of Interest. 
Figure~\ref{mgard-pipeline} shows the MGARD compression pipeline. MGRAD decorrelates data through a specially designed multilevel decomposition process. Specifically, the data is treated as a piecewise linear function $u$ that takes the same values as the original data for each node, and it is recursively decomposed level by level. 
The function approximation $Q_lu$ on level $l$ is projected onto the next level $l-1$ by (1) calculating multi-level coefficients $mc_l = I-\Pi_{l-1})Q_{l}u$, which is the difference between the values of the fine grid nodes at level $l$ and their corresponding piecewise linear approximations; and (2) calculating and applying the global correction $Q_{l-1}u=Q_lu+\Omega_lmc_l$.
Then, the multi-level coefficients can be compressed and used to reconstruct the approximation of the original data.

Algorithm~\ref{mgard-imp} shows the implementation of MGARD using \ours.
At each level, $mc_l$ is first computed using multi-linear interpolation (\texttt{lerp}) that can be done using our locality abstraction (line 6).  Line 7-9 calculate the global correction that applies $L^2$ projection of $mc_l$ to $u$. This process involves a transfer-mass matrix multiplication that can be done using the locality abstraction and a tri-diagonal linear system solver, which requires iterative abstraction since calculations in solving each system are sequential. Line 10 applies the correction followed by multilevel coefficients collected in line 11.
After all levels have been decomposed, linear quantization is applied to the coefficients with different quantization bin sizes applied to different levels to improve the compression ratio and capability to preserve the quantities of interest. We use map and process abstraction to first map multilevel coefficients to different levels and then apply different quantization functions.
Finally, we use entropy encoding, such as Huffman compression, to compress quantized data.

\SetKwInOut{KwInOut}{In/Out}
\SetKwInOut{KwIn}{In}
\SetKwInOut{KwOut}{Out}
\begin{algorithm}[ht!]
\caption{Huffman Data Compression}
\label{huffmen-imp}
\SetKwFunction{FMain}{Huffman}
\SetKwProg{Fn}{Function}{:}{\KwRet{$u_{compressed}$}}
\Fn{\FMain{}}{
\KwIn{Data $u$}
$freq \leftarrow$ \texttt{Global(u, histogram())}\\
\texttt{SortByKey(freq)}\\
\texttt{Global(freq, filter\_non\_zeros())}\\
$codebook \leftarrow$ \texttt{Global(freq, tree\_build())}\\
$u_{enc} \leftarrow $ \texttt{Locality(u, encode(codebook))}\\
$u_{compressed} \leftarrow $ \texttt{Global($u_{enc}$, serialize())}\\
}
\end{algorithm}

\subsection{Huffman lossless compression}

\begin{figure}[h!]
    \centering
    \includegraphics[width=0.5\textwidth]{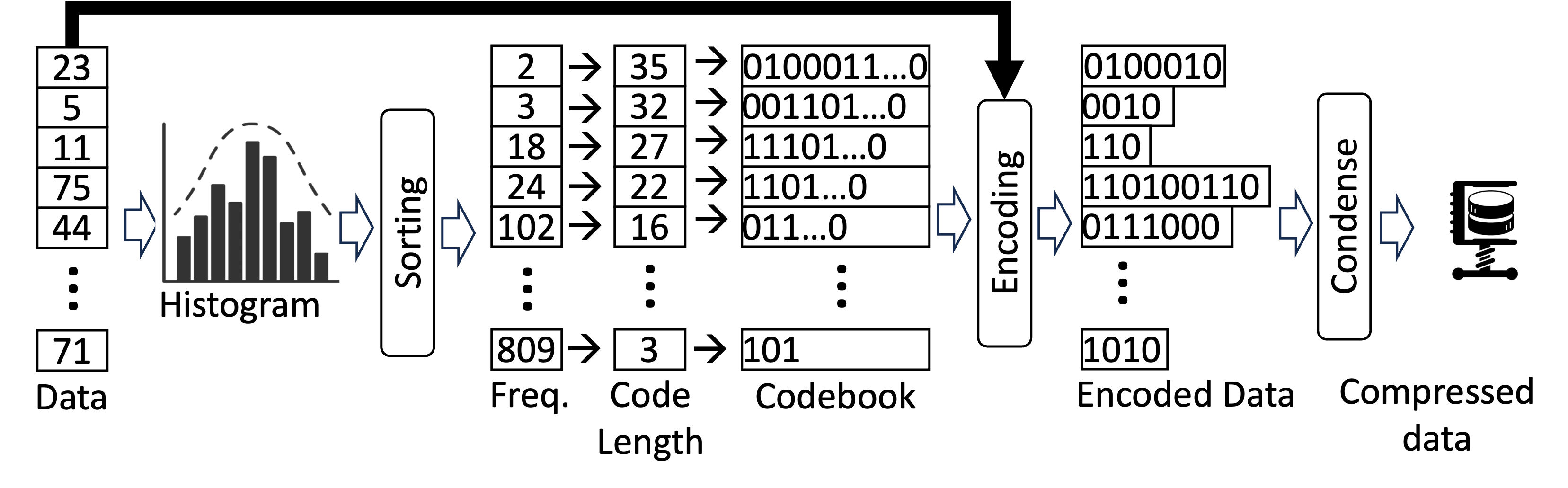}
    \caption{Huffman compression pipeline}
    \label{huffman-pipeline}
    \vspace*{-1em}
\end{figure}
Huffman is a widely used lossless compression in many scientific data compressors. It works as an entropy encoder that can effectively compress decorrelated data. Figure~\ref{huffman-pipeline} shows the Huffman compression pipeline. It works by first collecting the frequency distribution of input elements using a histogram, then using the frequencies to generate variable-length Huffman codes for different keys. The encoding process replaces the original values in data with Huffman codes such that more frequent keys are represented with fewer bits. Finally, encoded variable-length keys are condensed into a compact format that forms the compressed data.

Algorithm~\ref{huffmen-imp} shows the implementation of Huffman using \ours.
In line 2, we first collect the frequency of each key using the histogram operation proposed in~\cite{gomez2013optimized}. 
As all threads need to collaboratively update the frequency counters, we use the global abstraction.
After sorting the frequencies in line 2, we filter all the keys that have non-zero frequency in line 3.
Next, at line 5, we implement the parallel two-phase Huffman treeless codebook generation algorithm~\cite{accelerated2007two} as it provides much higher parallelism.
Limited by the page space, we refer readers to~\cite{tian2021revisiting} for a detailed algorithm for codebook generation.
At line 6, we encode the input key using the codebook. As each key can be encoded independently, we use the locality pipeline abstraction to exploit maximum parallelism.
Finally, we serialize all encoded keys into a compact form. Since all threads need to write to the same buffer, global coordination is required to avoid conflicts.

\SetKwInOut{KwInOut}{In/Out}
\SetKwInOut{KwIn}{In}
\SetKwInOut{KwOut}{Out}
\begin{algorithm}[ht!]
\caption{ZFP Fix-Rate Compression}
\label{zfp-imp}
\SetKwFunction{FMain}{ZFP}
\SetKwProg{Fn}{Function}{:}{\KwRet{$u_{compressed}$}}
\Fn{\FMain{}}{
\KwIn{Data $u$}
\KwIn{Bit rate $r$}
$u_{aligned} \leftarrow$ \texttt{Locality(u, exp\_align())}\\
$u'\leftarrow$ \texttt{Locality(u, orthogonal\_transform(u))}\\
$u_{compressed} \leftarrow$ \texttt{Locality(u', bitplane\_encode(r))}\\
}
\end{algorithm}

\subsection{ZFP fix-rate compression}

\begin{figure}[h!]
    \centering
    \includegraphics[width=0.5\textwidth]{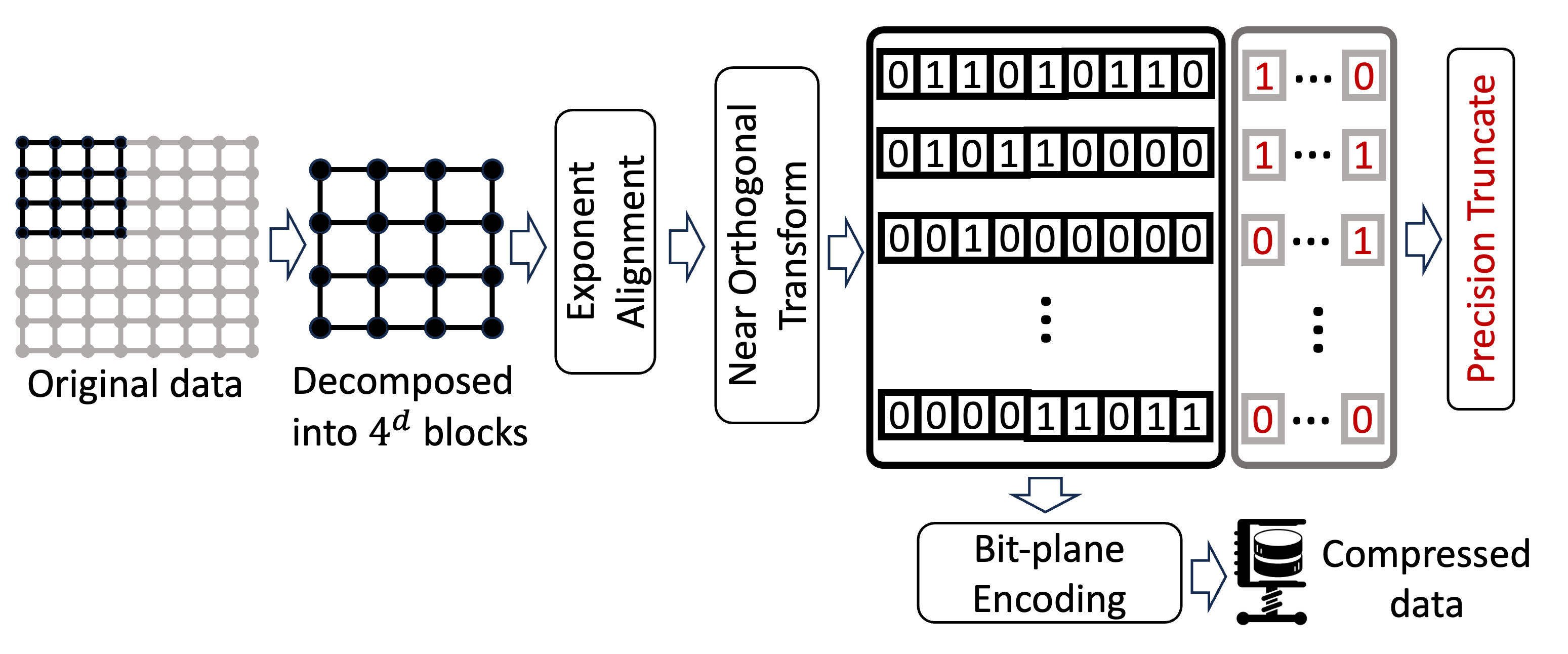}
    \caption{ZFP compression pipeline}
    \label{zfp-pipeline}
    \vspace*{-1em}
\end{figure}

ZFP is designed to enable high throughput compression as it brings relatively lower computational costs compared with other compression pipelines. 
Figure~\ref{zfp-pipeline} shows the ZFP compression pipeline. The input data is first decomposed into blocks of $4^d$ (d: number of dimensions). The exponents of elements of each block is first aligned according to the max components and converted to fixed-point values. Then those values are transformed using a customized near-orthogonal transformation to be more compressible. Then a number of less significant bits of transformed coefficients are truncated and the rest of the bits are serialized into bitplanes that forms compression data stream.

Algorithm~\ref{zfp-imp} shows the implementation of ZFP fix-rate compression using \ours.
ZFP has three compression modes: fix-rate, fix-accuracy, fix-precision.
To demonstrate the applicability of \ours, we choose to only implement the fix-rate mode. The other two modes can be implemented in similarly.
Since both exponent alignment and near-orthogonal transformation operations are applied blockwise, we can implement them using the locality abstraction (line 2-3).
The transformed data needs to be truncated and serialized (line 4) to form the compression data. As all blocks output the same size bit streams, this can be done without global coordination, we also apply locality abstraction.

\section{Pipeline optimization}
\label{sec:pipeline-opt}

To minimize the data movement overhead, especially serious in GPU-based data reduction pipelines, we build a carefully tailored pipeline optimization in \ours for data reduction.
Although we mainly target GPU-based pipelines, our optimization can be portable across different processors.

\begin{figure}[h!]
\centering
\includegraphics[width=0.5\textwidth]{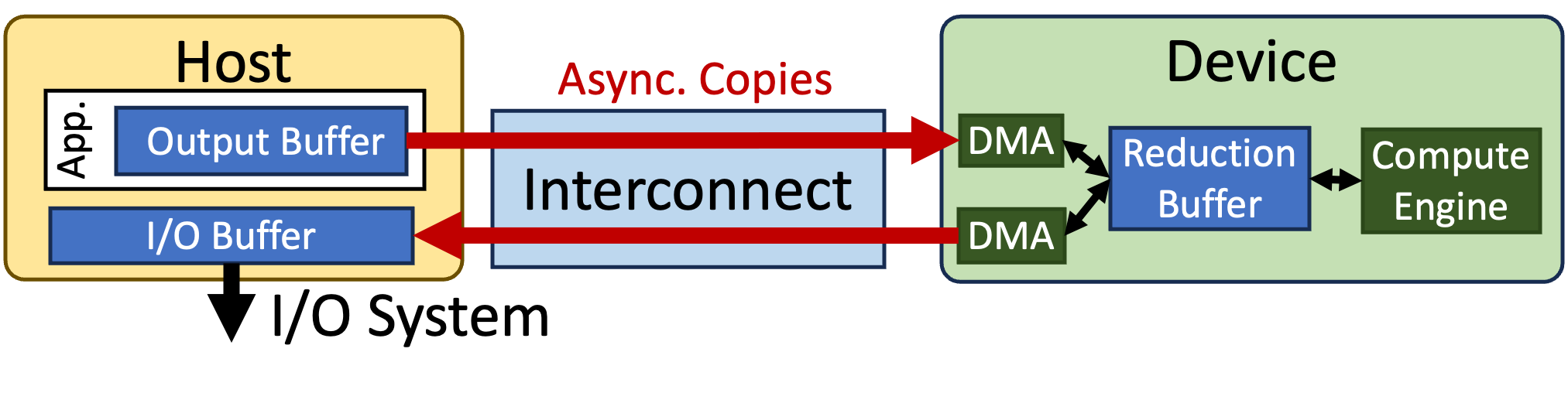}
\vspace*{-2em}
\caption{Host-Device Execution Model}
\label{machine-abs}
\vspace*{-1em}
\end{figure}

\subsection{Host-Device Execution Model}
First, to enable portability across different GPU-accelerated systems, we use a machine abstraction, Host-Device Execution Model (HDEM), to design our pipeline. Figure~\ref{machine-abs} shows our execution model that represents the common architecture in many modern GPU-based systems. One GPU device in our abstraction has two Direct Memory Access (DMA) engines, each of which can work independently for asynchronous memory copy. They can support data copies between the application buffer, I/O buffer, and reduction buffer. The device also has a compute engine to support the concurrent execution of reduction kernels during data copy operations.

\begin{figure}[ht]
    \centering
    \includegraphics[width=0.5\textwidth]{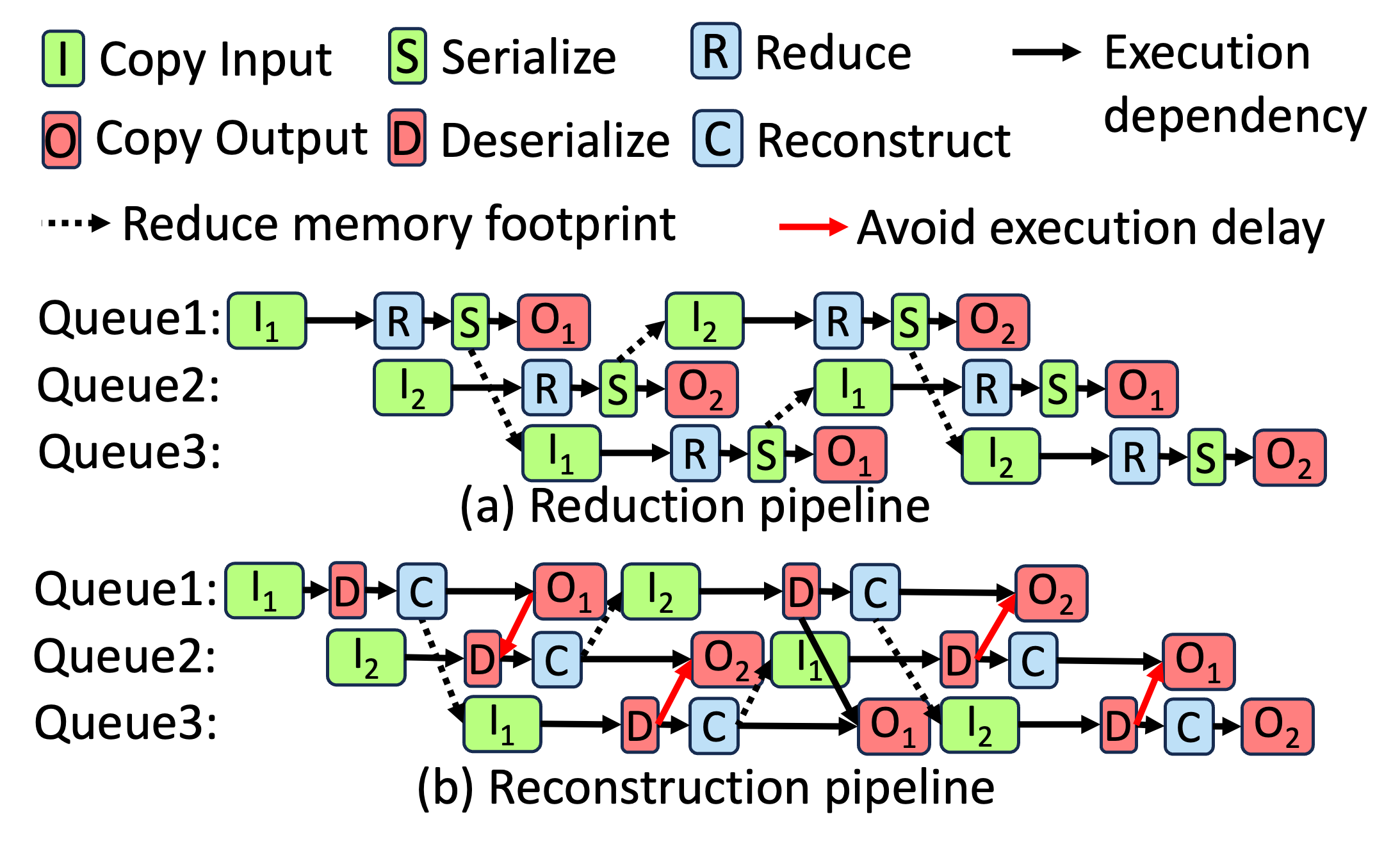}
    \vspace*{-1em}
    \caption{Optimized reduction and reconstruction pipeline represented as DAGs that enable data transfer latency hiding. Green: host-to-device copy; Red: device-to-host copy; Blue: compute}
    \label{opt-pipeline}
    \vspace*{-1em}
\end{figure}

\subsection{Optimized Pipeline}
Next, to ease the design of our pipeline, we consider two restrictions: (1) we assume reduction kernels are already highly optimized by developers so that they can achieve maximum occupancy with large input sizes. So, we restrict only one kernel running at the same time. 
(2) As moving both input and output data can take a considerable amount of time in a reduction pipeline, we dedicate one DMA for moving input data and another for moving output to exploit data movement overlap in opposite directions. We do not consider the data movement overlaps in the same direction (e.g., copying output and deserialization in the reconstruction pipeline) since the potential improvement is limited.

To fully use the hardware throughput, our pipeline needs to have enough depth for latency hiding. Depending on the time complexity of the reduction algorithm, data compressibility, and hardware performance, the overall performance of our pipeline could rely on the throughput of one of the DMAs or the compute engine. So, according to Little's law~\cite{little2008little}, three is the minimum depth for our pipeline to fully use the hardware.
Figure~\ref{opt-pipeline} shows our optimized pipeline for both reduction and reconstruction.
In the figure, the reduction process is pipelined among three queues (1-3). Green boxes represent host-to-device (H2D) DMA copy tasks. Red boxes represent device-to-host (D2H) DMA copy tasks. Blue boxes are compute tasks. According to our restrictions, no two tasks with the same color can be executed at the same time. We assume serialization and deserialization are needed for embedding and extracting metadata after and before computation, which also relies on D2H and H2D copies. 

Theoretically, a pipeline with three execution queues requires three distinct input/output buffers for correctness. This imposes challenges for data reduction since a high memory footprint could limit the input data chunk size of each reduction process, which in turn limits the overall data compressibility and computing efficiency of reduction. To reduce the memory footprint, we insert additional dependencies in the pipeline (marked as dotted lines) to avoid data races between executions on queue $X$ and $(X+2)\%3$. This optimization reduces the needed input/output buffers to two: $I_1$/$O_1$ and $I_2$/$O_2$.
To explain how the extra dependencies eliminate the needs for the third set of buffers, we use the dependency between $I_1$ and $S$ in the reduction pipeline as an example. 
This dependency enforces that the third reduction pipeline must not start until the serialization of the first reduction process finishes, which indicates the buffer $I_1$ is no longer needed. So, we can immediately reuse $I_1$ as the input buffer for the third reduction process;
We add the same dependencies to the reconstruction pipeline. 
In addition, we also add dependencies (red arrows) to optimize launching orders for each operation. During the reconstruction process, both the deserialization operation of the next upcoming process and the output copy operation rely on the same D2H DMA, which will cause serialized execution. By default, copying output is initialized right after the previous reconstruction and before the deserialization of the next reconstruction. However, this can cause serious delays to the next reconstruction. So, we reverse the order of the two such that deserialization is done first, and the reconstruction overlaps with the output copy operation.
\begin{figure}[h!]
    \centering
    \includegraphics[width=0.5\textwidth]{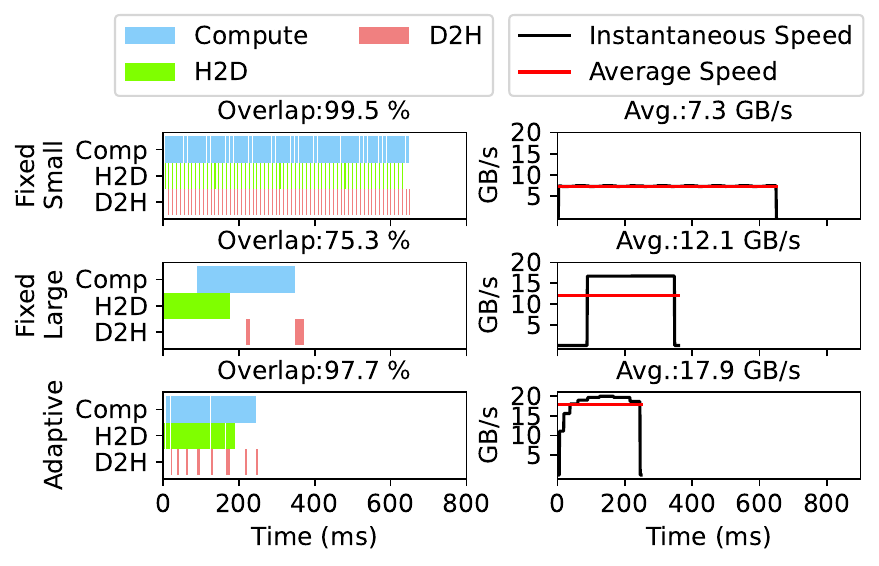}
    \vspace*{-1em}
    \caption{
    Reduction pipeline of compressing a 4.3 GB NYX variable with $1e^{-2}$ error bound using MGARD. Fixed small and large use fixed chunks of size 100 MB and 2 GB. Adaptive represents our adaptive chunk size adjustment strategy. 
    }
    \label{pipeline-nyx}
\end{figure}

\SetKwInOut{KwInOut}{In/Out}
\SetKwInOut{KwIn}{In}
\SetKwInOut{KwOut}{Out}
\begin{algorithm}[h!]
\caption{Adaptive Pipeline}
\label{adaptive-chunk}
\SetKwFunction{FMain}{AdaptivePipeline}
\SetKwProg{Fn}{Function}{:}{\KwRet{$u_{reduced}$}}
\Fn{\FMain{}}{
\KwIn{Data $u$} 
\KwIn{Initial chunk size $C_{init}$} 

$C_{curr} \leftarrow C_{init}$;

$in\_offset, out\_offset \leftarrow 0$

$size_{rest} \leftarrow LargestDim(u)$

$Q[3] \leftarrow$ init. device queues

$I[2] \leftarrow$ init. input device buffers

$O[2] \leftarrow$ init. output device buffers

$i, j \leftarrow$ 0
$C_{limit} \leftarrow$ maximum chunk size

$I[j]\xleftarrow[copy\ size = C_{curr}]{H2D\ on\ queue\ Q[i]} u[offset]$

$size_{rest} = size_{rest} - C_{curr}$

$in\_offset = in\_offset + C_{curr}$

\While{$size_{rest} \geq 0$}{

$i_{next} = (i + 1) \% 2$

$j_{next} = (j + 1) \% 3$

\If{$size_{rest} > 0$}{

$C_{next} \leftarrow min(\Theta(C_{curr}/\Phi(C_{curr})), C_{limit})$

$C_{next} \leftarrow min(size_{rest}, C_{next})$

$I[j_{next}]\xleftarrow[copy\ size = C_{next}]{H2D\ on\ queue\ Q[i_{next}]} u[offset]$

$size_{rest} = size_{rest} - C_{curr}$

$in\_offset = in\_offset + C_{curr}$

}

$O[j]\xleftarrow{Reduce\ on\ queue\ Q[i]} I[j]$

$u_{reduced}[out\_offset]\xleftarrow[]{D2H\ on\ queue\ Q[i]} O[j]$

$out\_offset = out\_offset + O[j].size$\\
$C_{curr} = C_{next}$

$i = i_{next}$

$j = j_{next}$

}
}
\end{algorithm}
\subsection{Optimizing chunk size}
The chunk size can greatly impact the overall pipeline performance in terms of compute efficiency of reduction kernels on GPU and the effectiveness of compute and communication overlap. We define the overlap ratio as:
$$Overlap = \frac{\text{Total overlapped H2D and D2H time}}{\text{Total H2D and D2H time}}$$
On one hand, when the chunk size is small, the pipeline tends to have a high overlap ratio because reduction computation can start soon after the first small chunk is transferred to the GPU. However, due to limited input size, the computing efficiency may be limited.  On the other hand, when the chunk size is large, each chunk is large enough to fully utilize the GPU to achieve high reduction throughput. However, the reduction process needs to wait for a longer time for the first chunk to be transferred. Figure~\ref{pipeline-nyx} shows the effect of using different chunk sizes. With a small chunk size, the sustained throughput is low (7.3 GB/s) as chunks are too small to fully occupy the GPU. When the chunk size is large, only 75.3\% of the data transfer latency can be hidden.
\begin{figure}[h]
\centering
\includegraphics[width=0.5\textwidth]{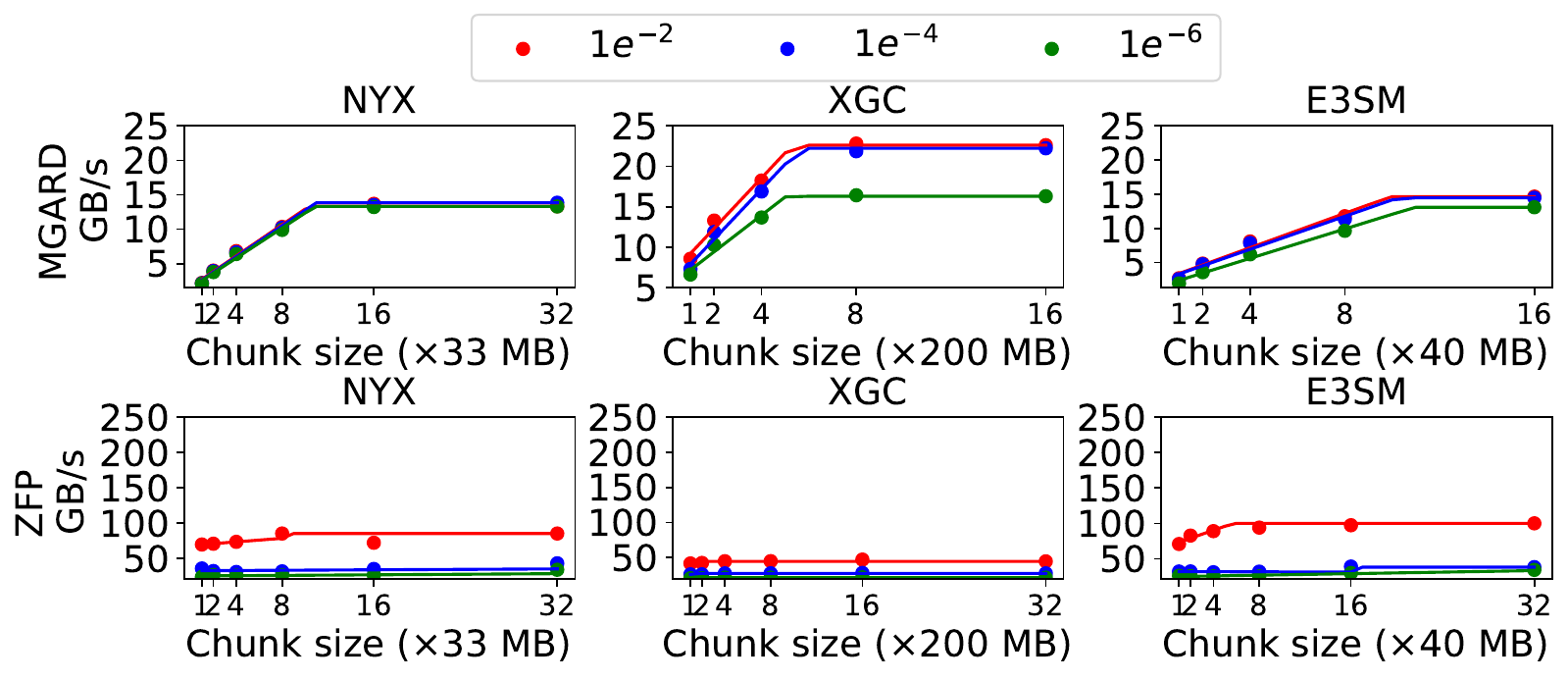}
\caption{Modeling MGARD and ZFP performance of different chunk sizes using the Roofline model}
\label{model-speed}
\end{figure}
To both improve the overlap ratio while maximizing sustained throughput, we build a reduction pipeline with an adaptive chunk size adjusting strategy as shown in
Algorithm~\ref{adaptive-chunk}.
Our algorithm first uses a small user-specified chunk size as the initial size (line 2) to reduce the leading time of the whole pipeline.
To minimize the GPU computing inefficiency caused by small chunk sizes, we adaptively increase the chunk size as reduction make progresses.
We determine the next chunk size by estimating how much data we can transfer to GPU while GPU is working on the current chunk.
This can ensure the memory copy time is completely hidden by the compute time.
This needs us to build two estimation functions: (1) $p = \Phi(C)$: estimating the reduction throughput $p$ on GPU given a chunk size $C$ (2) $C_{max} = \Theta(t)$: estimating the maximum chunk size $C_{max}$ can be transferred from host to device given a time limit $t$. Then, we can use the current chunk size $C_{curr}$ and the maximum chunk size limited by GPU memory $C_{limit}$ to compute the next chunk size $C_{next} = min(\Theta(C_{curr}/\Phi(C_{curr})), C_{limit})$ (line 17).
We build $\Phi(C)$ using our modified roofline model:
$$\Phi(C) = \begin{cases} \alpha \times C + \beta
 & \text{ if } C < C_{threshold} \\ 
\gamma & \text{ if } C \geq C_{threshold} 
\end{cases}$$
We use a linear model while chunk sizes are small and GPU is not saturated and use a constant function to estimate saturated throughput when chunk size is larger than a threshold $C_{threshold}$.
The model can be obtained by profiling a given dataset and error bound on different chunk sizes until the chunk is large enough to saturate the GPU.
We use the throughput of the largest profiled chunk to determine $\gamma$ and orderly check on the throughput of smaller chunks until the throughput drops below $f \times \gamma$ (e.g., $f=0.1$), then we use linear regression to fit the rest chunks sizes to obtain the linear model.
Figure~\ref{model-speed} shows the profiling results and fitted models on three datasets with three error bounds.
As for $\Theta$ we use a linear model to estimate the maximum transferable chunk size:
$\Theta(t) = \frac{t}{\beta}$.
We treat host-to-device throughput $\beta$ as constant since we do not consider small chunk sizes that do not saturate CPU-GPU interconnect, which would lead to an inefficient pipeline.




\begin{figure}[t]
\centering
\includegraphics[width=0.5\textwidth]{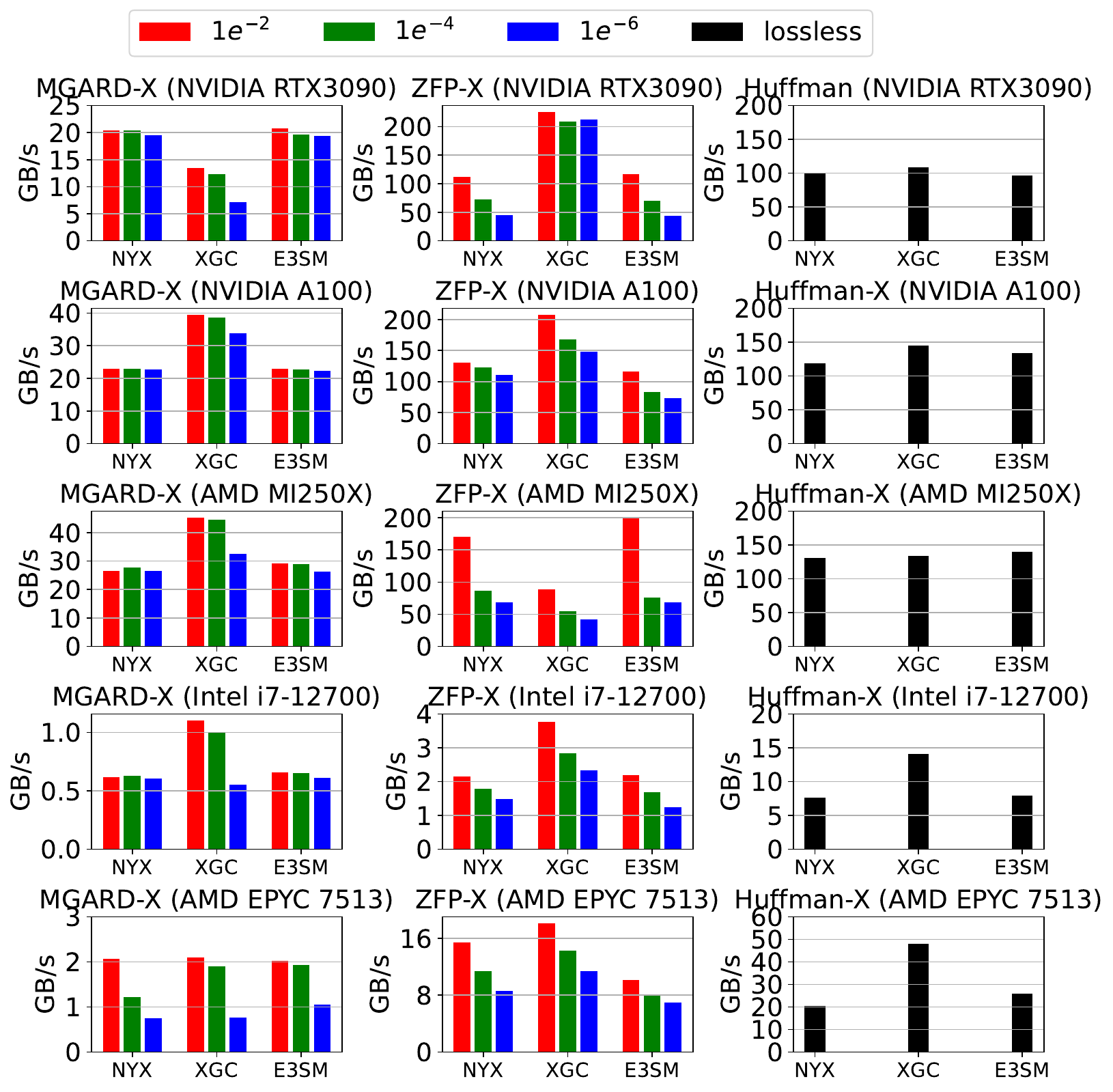}
\vspace*{-0.5em}
\caption{Kernel throughput of portable MGARD-X, ZFP-X, and Huffman-X implementations on five processors. MGARD-X and ZFP-X compress with three different relative error bounds and Huffman-X provides lossless compression.}
\label{portability}
\vspace*{-1em}
\end{figure}

\section{Experimental Evaluation}
\label{sec:eval}
\subsection{Experimental Methodology}
We implemented three data reduction pipelines, MGARD~\cite{gong2023mgard}, ZFP~\cite{lindstrom2014fixed}, and Huffman~\cite{tian2021revisiting}, using \ours based on their published algorithm designs.
We name our portability implementation as MGARD-X, ZFP-X, and Huffman-X.
As comparison reference, we use the current release version of GPU implementation of MGARD, ZFP, SZ, and LZ4: \texttt{MGARD-GPU v1.5}~\cite{chen2021accelerating}, \texttt{ZFP-CUDA v1.0}~\cite{lindstromzfp}, \texttt{cuSZ v0.6}~\cite{tian2021cusz}, \texttt{NVCOMP-LZ4 v2.2}~\cite{nvcomp}.
At the time of evaluation, ZFP only supports fix-rate mode on GPU, so we only use fix-rate mode in our evaluation.
Also, we restrict our evaluation to only compare with stable version of each software, so we exclude the evaluation of HIP version of SZ and ZFP on Frontier.
For parallel I/O evaluations, we integrate each reduction routine into the I/O pipeline of the ADIOS 2 library~\cite{godoy2020adios} and use the latest BP5 file format. Aggregation strategy is tuned for each system (i.e., one writer per node on Summit and one writer per GPU on Frontier) to achieve the best I/O performance. 
In addition, three scientific datasets are used in our data reduction evaluations:

\begin{table}[h!]
\caption{Datasets used for evaluation}
\begin{tabular}{|c|c|c|c|c|}
\hline
Dataset     & Field & Dimensions & Data Type & Size  \\ \hline
NYX & density &$512 \times 512 \times 512$        & FP32        & 536.8MB \\ \hline
XGC       &e\_f & $8 \times 33 \times 1117528 \times 37$        & FP64        & 87.3 GB \\ \hline
E3SM  & PSL & $2880 \times 240 \times 960$        & FP32        & 2.7 GB \\ \hline
\end{tabular}
\vspace*{-1em}
\end{table}

\begin{figure}[t]
\centering
\includegraphics[width=0.5\textwidth]{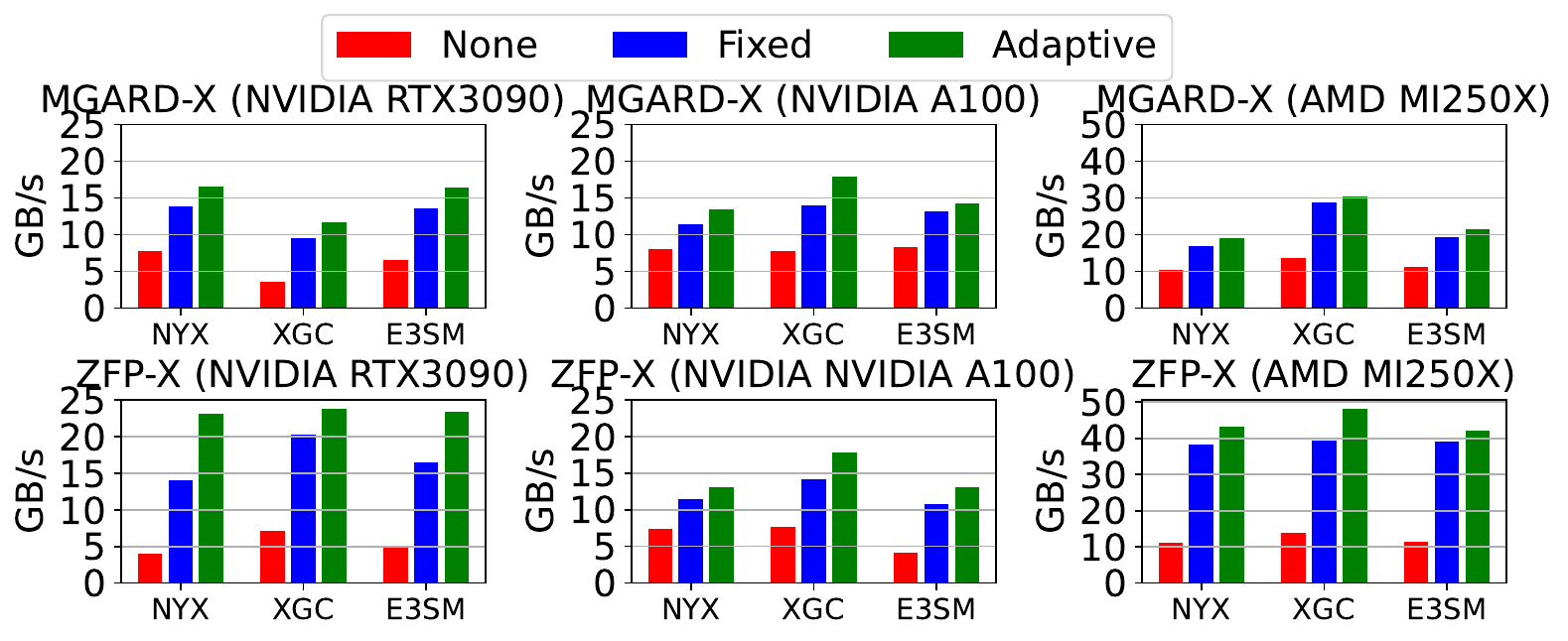}
\vspace*{-0.5em}
\caption{Comparing end-to-end throughput of MGARD and ZFP using (1) None: no overlapping pipeline; (2) Fixed: fixed size pipeline; and (3) Adaptive: adaptive size pipeline}
\label{singe-pipeline}
\vspace*{-1em}
\end{figure}

\begin{figure}[t]
\centering
\includegraphics[width=0.5\textwidth]{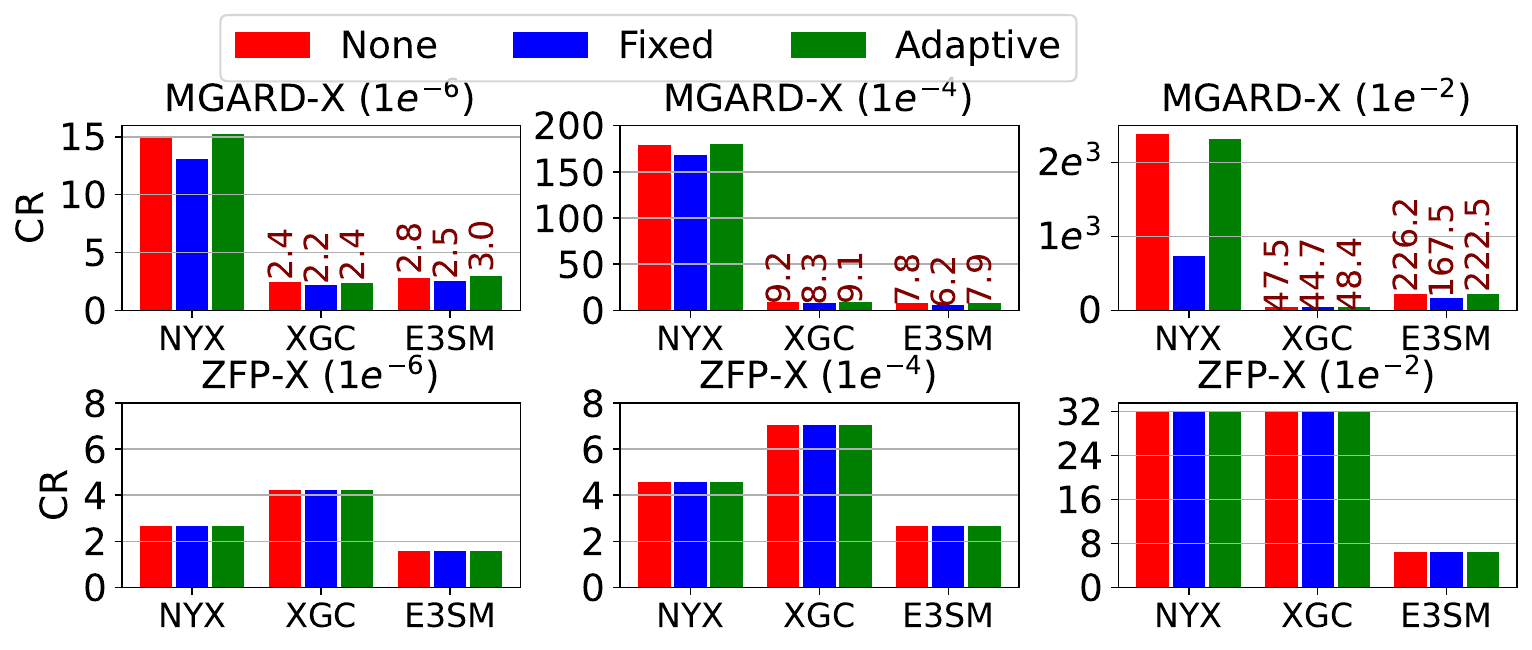}
\vspace*{-0.5em}
\caption{Comparing compression ratio of MGARD and ZFP with three pipeline settings and errors: $1e^{-2}$, $1e^{-4}$, and $1e^{-6}$}
\label{singe-pipeline-cr}
\vspace*{-1em}
\end{figure}

\subsection{Experimental Environment}
All evaluates are done on four platforms: Summit, Frontier, Jetstream2, and a GPU workstation. Summit and Frontier are two leadership class supercomputers at Oak Ridge Leadership Computing Facility (OLCF).
Summit is a 4,608-node pre-exascale supercomputer using a GPFS filesystem with peak I/O bandwidth of 2.5 TB/s.
Each compute node is equipped with 6 NVIDIA V100 GPUs 
with 16 GB memory on each GPU and two 22-core IBM POWER9 CPUs with 512 GB memory. 
Frontier is a 9,408-node exascale supercomputer using a Luster filesystem with peak I/O bandwidth of 9.4 TB/s.
Each compute node is equipped with 4 AMD Instinct MI250X GPUs with 128 GB memory on each GPU and one 64-core AMD EPYC CPU with 512 GB memory.
The Jetstream2 is a system provided by the Indiana University via the NSF ACCESS program~\cite{boerner2023access}.
Jetstream2 includes 90 GPU-enabled nodes and each node is equipped with 4 NVIDIA A100 GPUs with 40 GB memory on each GPU and two 64-core AMD Milan 7713 CPUs with 512 GB memory.
The GPU workstation is equipped with an NVIDIA RTX 3090 GPU with 24 GB memory and 20-core Intel i7 CPU with 32 GB memory. 

\subsection{Kernel-level portability and performance}
We evaluate the portability and performance of \ours on five different CPU and GPU processors for each reduction kernel we implemented.
Figure~\ref{portability} shows the performance of reduction kernel excluding the data transferring cost between host and device.
On GPUs, we achieve up to 45 GB/s, 210 GB/s, and 150 GB/s throughput for MGARD-X, ZFP-X, and Huffman-X respectively. On CPUs, we obtain up to 2 GB/s, 18 GB/s and 48 GB/s throughput for the three reduction kernels.

\begin{figure}[t]
\centering
\includegraphics[width=0.5\textwidth]{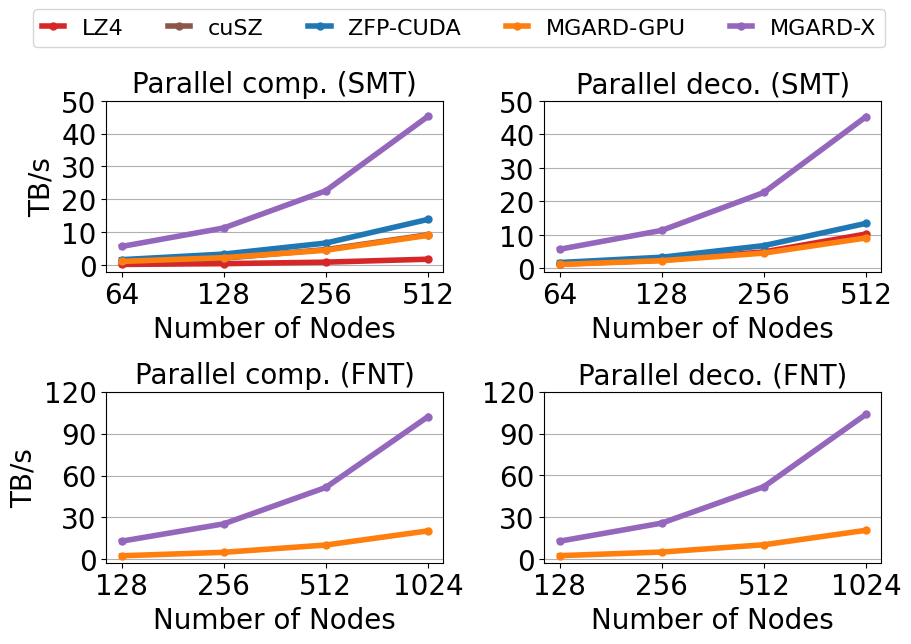}
\caption{Aggregated compression and decompression throughput on Summit (SMT) and Frontier (FNT)}
\label{scale-multi-node}
\vspace*{-1em}
\end{figure}

\subsection{Single-GPU end-to-end pipeline performance}
Next, we evaluate the end-to-end pipeline performance on a single GPU with and without our proposed optimization techniques.
We exclude applying pipeline optimization on CPUs because the data transfer time typically is not the dominating factor of the whole reduction pipeline such that our optimizations can only achieve limited improvement.
Figure~\ref{singe-pipeline} shows the end-to-end pipeline performance including both data transfer and computation costs. Compared with not using an overlapping pipeline, our fixed size pipeline (100 MB chunk) achieved up to $2.1\times$ and $3.5\times$ speedups for MGARD-X and ZFP-X, respectively.
In addition, compared with the fixed size pipeline, our adaptive size pipeline achieved up to $1.3\times$ and $1.6\times$ speedups for the three reduction techniques, respectively.
Figure~\ref{singe-pipeline-cr} compares the compression ratios when using three different pipeline settings. Compared with not using overlapping pipelines, the fixed size pipeline reduces the compression ratios by $5-67\%$ for MGARD due to degraded compressibility of the small chunk size. Our Adaptive pipeline brings a similar compression ratio ($<1\% $ difference) compared with the non-pipeline setting, as it leverages large chunk sizes that have better compressibility. The pipeline setting brings negligible impact to ZFP as it compresses data blocks by blocks with much smaller generality than our chunk size.

\begin{figure}[t]
\centering
\includegraphics[width=0.5\textwidth]{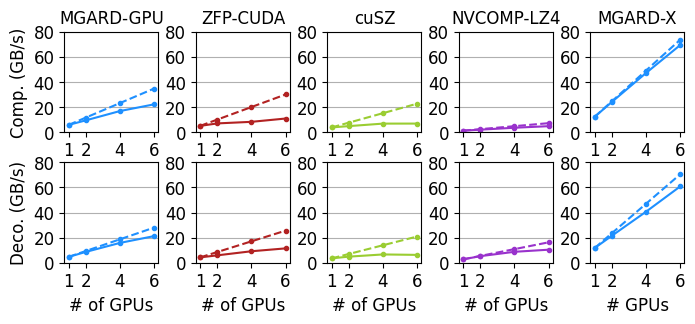}
\vspace*{-1em}
\caption{Scalability on multiple V100 GPUs. Solid: real throughput; Dotted: throughput of ideal scalability}
\label{scale-1-node}
\vspace*{-1em}
\end{figure}

\subsection{Multi-GPU end-to-end performance and scalability}
As modern supercomputers commonly use dense multi-GPU architectures, we evaluate the performance and scalability of \ours on a Summit node with 6 V100 GPUs.
As shown in Figure~\ref{scale-1-node}, we use MGARD-X as an example to compare the performance and scalability of \ours with four other data reduction not implemented using \ours.
To measure the scalability, we calculate the ideal speed at scale by multiplying the speed on one GPU with the number of GPUs.
To quantify the scalability, we calculate the average real-to-ideal speed ratios across tests on different numbers of GPUs.
Because of our context memory management optimization, \ours minimizes most of the device memory operations during the execution of the pipelines, which could cause contention in the multi-device environment.
For compression, MGARD-X achieved 96\% avg. scalability while the non-optimized MGRAD, ZFP, SZ, and LZ4 achieved 72\%, 48\%, 46\%, and 74\% avg. scalability.
Similarly for decompression, MGARD-X achieved 88\% avg. scalability while existing designs achieved 76\%, 55\%, 48\%, and 70\% avg. scalability.

\subsection{Multi-node end-to-end performance}
Figure~\ref{scale-multi-node} shows the multi-node aggregated data reduction end-to-end throughput on Summit and Frontier. We perform a weak scaling test where each node compression and decompresses the NYX data. 
To fully saturate the data reduction pipelines on each GPUs, we let each GPU process 14 time steps of NYX data.
For Summit, we scale the compression and decompression process up to 512 nodes using the computing power of 3,072 V100 GPUs and 23 TB of data. 
At this scale, MGARD-X achieved 45 TB/s throughput while NVCOMP-LZ4, cuSZ, ZFP-CUDA, and MGARD-GPU only achieved 10 TB/s, 9 TB/s, 13 TB/s, and 9 TB/s throughput respectively.
For Frontier, we scaled our test using up to 1,024 nodes using the computing power of 4,096 MI250X GPUs and 62 TB of data. At this scale, MGARD-X achieved 103 TB/s throughput while MGARD-GPU only achieved 18 TB/s.

\subsection{Weak scaling I/O acceleration evaluation}
\begin{figure}[t]
\centering
\includegraphics[width=0.5\textwidth]{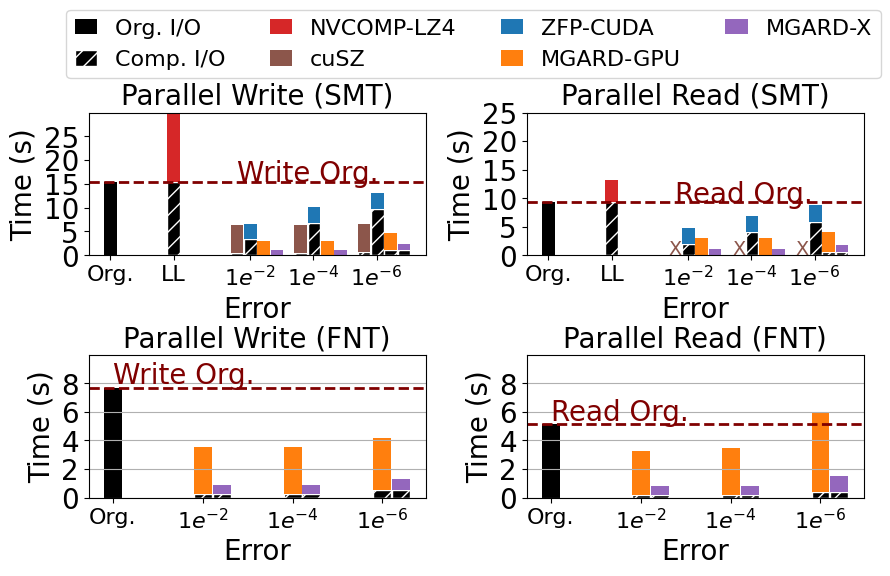}
\vspace*{-1em}
\caption{Weak scaling parallel I/O using NYX data on Summit (SMT) and Frontier (FNT)}
\label{weak-scale-io-nyx}
\vspace*{-1em}
\end{figure}

To evaluate the effectiveness of using data reduction to accelerate parallel I/O at scale, we first perform weak scaling I/O using the NYX data on Summit with 512 nodes and Frontier with 1,024 nodes. Each GPU compresses 7.5 GB of data. 
Figure~\ref{weak-scale-io-nyx}(a) shows the results on Summit. Compared with the original read and write cost, NVCOMP-LZ4 cannot bring I/O acceleration due to a limited compression ratio ($1.1\times$) with extra computational overhead, which leads to an extra 83.5\% and 42.7\% read/write overhead. cuSZ achieved $2.3 - 2.4 \times$ write acceleration with compression ratio $20 - 31\times$.  However, cuSZ crashes at scales larger than 64, so we could not measure its read acceleration. 
ZFP-CUDA achieved $1.2-2.3\times$ write acceleration and $1.1-1.9\times$ read acceleration with $2.4-32\times$ compression ratio.
MGARD-GPU archived $3.3-5.1\times$ write acceleration and $2.3-3.1\times$ read acceleration with $14-2379\times$ compression ratio.
MGARD-X archived $6.8-15.3\times$ write acceleration and $5.2-9.3\times$ read acceleration with the same compression ratio as MGARD-GPU. 
Figure~\ref{weak-scale-io-nyx}(b) shows the results on Frontier. Compared with the original read and write cost, MGARD-GPU archived $1.8-2.1\times$ write acceleration and $0.8-1.5\times$ read acceleration with the same compression ratio as on Summit.
MGARD-X archived $6.0-8.5\times$ write acceleration and $3.5-6.5\times$ read acceleration with the same compression ratio as MGARD-GPU.


\subsection{Strong scaling I/O acceleration evaluation}
\begin{figure}[t]
\begin{subfigure}[t]{0.5\textwidth}
\centering
\includegraphics[width=\textwidth]{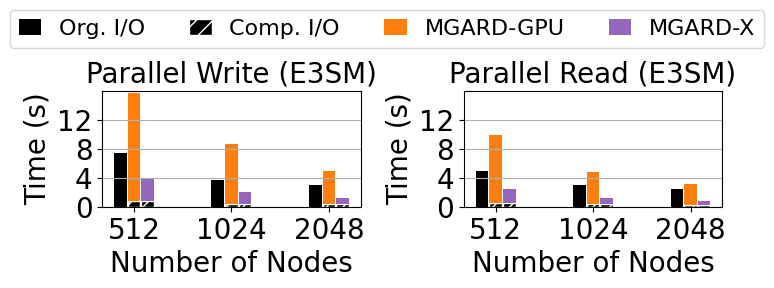}
\end{subfigure}
\begin{subfigure}[t]{0.48\textwidth}
\centering
\includegraphics[width=\textwidth]{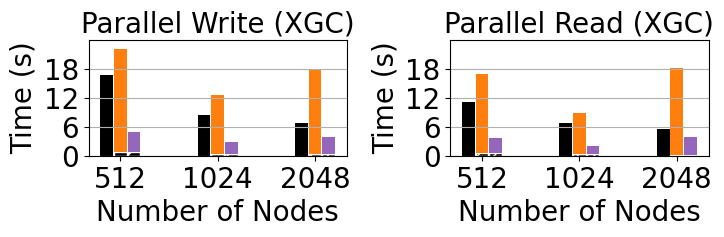}
\end{subfigure}
\vspace*{-1em}
\caption{Strong scaling parallel I/O on Frontier}
\label{strong-scale-io}
\vspace*{-1em}
\end{figure}
In addition to the weak scaling test, we also perform strong scaling write and read I/O test with and without data reduction on Frontier.
Figure~\ref{strong-scale-io} (a) shows the I/O cost of writing and reading 32 TB of E3SM data using 512, 1024, and 2048 nodes.
The data is compressed with a relative error bound of $10^{-4}$ with a compression ratio of $7.9\times$ for both MGARD-GPU and MGARD-X. Compared with I/O without reduction, MGARD-GPU brings 28\% - 134\% extra overhead due to low reduction throughput. MGARD-X, on the other hand, can accelerate write by $2.4-1.8\times$ and read by $2.1-2.9\times$ across different scales.
Figure~\ref{strong-scale-io} (a) shows the I/O cost of writing and reading 67 TB of XGC data using 512, 1024, and 2048 nodes.
The data is compressed with a relative error bound of $10^{-4}$ with a compression ratio of $9.1\times$ for both MGARD-GPU and MGARD-X. Compared with I/O without reduction, MGARD-GPU brings 32\% - 227\% extra overhead due to low reduction throughput. MGARD-X, on the other hand, accelerates write by $1.7-3.4\times$ and read by $1.5-3.3\times$ across different scales.



\section{Conclusion}
The increasing gap between scientific data generation and the capabilities of computing systems to process and analyze this data underscores the urgent need for effective data reduction strategies. While GPU-accelerated data reduction techniques have shown promise, several challenges, such as limited portability, memory-bound performance, and scalability issues, still hinder their broader adoption in exascale workflows. To address these obstacles, we developed \ours, a portable, high-performance data reduction framework that optimizes memory transfer overhead and enhances scalability across multiple GPU and CPU architectures. Through rigorous testing and integration with large-scale systems, \ours demonstrates significant improvements in data reduction throughput and I/O acceleration, highlighting its potential to transform scientific workflows in the exascale era.
\section*{Acknowledgment}
This work used Jetstream2 system through allocation CIS230203 from the Advanced Cyberinfrastructure Coordination Ecosystem: Services \& Support (ACCESS) program, which is supported by U.S. National Science Foundation (NSF) grants 2138259, 2138286, 2138307, 2137603, and 2138296.
This work was partially supported by the NSF under Grant OAC-2311756, OAC-2313122, and OIA-2327266.
This research was also supported by the SIRIUS-2 ASCR research project, the Scientific Discovery through Advanced Computing (SciDAC) program, specifically the RAPIDS-2 SciDAC institute. 
\bibliographystyle{IEEEtran}
\bibliography{ref}

\end{document}